\documentclass[aps, preprint, onecolumn,secnumarabic,nobalancelastpage,nofootinbib]{revtex4-1} 
\usepackage{graphicx}
\usepackage{amsmath,amssymb}
\usepackage[utf8]{inputenc}
\usepackage[T1]{fontenc}
\usepackage{lmodern}
\usepackage{natbib}
\usepackage{hyperref}
\usepackage{xcolor}
\hypersetup{colorlinks=true,citecolor={blue},linkcolor={blue},urlcolor={blue}}
\usepackage{units}
\usepackage[english]{babel}
\usepackage[normalem]{ulem}
\usepackage{upgreek}
\usepackage{wasysym}
\usepackage{placeins} 
\usepackage{soul}
\makeindex

\begin{document}

\definecolor{orange}{RGB}{255, 69, 0}
\definecolor{green}{RGB}{26,148,49}
\newcommand{\Lucy}[1]{{\color{green}{#1}}}
\setstcolor{red}
\newcommand{\hs}[1]{\textcolor{red}{#1}}
\newcommand{\jdt}[1]{\textcolor{orange}{#1}}

\title{Time-Delay Polaritonics}

\date{\today}

\author{J. D. T\"opfer$^{1,2*}$} 
\author{H. Sigurdsson$^{1,2}$}
\author{L. Pickup$^{2}$}
\author{P. G. Lagoudakis$^{1,2*}$}
\affiliation{$^1$Skolkovo Institute of Science and Technology Novaya St., 100, Skolkovo 143025, Russian Federation}
\affiliation{$^2$Department of Physics and  Astronomy, University of Southampton, Southampton, SO17 1BJ, United Kingdom}
\affiliation{$^*$email: J.D.Toepfer@soton.ac.uk, Pavlos.Lagoudakis@soton.ac.uk}

\renewcommand{\abstractname}{} 
\begin{abstract}
Non-linearity and finite signal propagation speeds are omnipresent in nature, technologies, and real-world problems, where efficient ways of describing and predicting the effects of these elements are in high demand. Advances in engineering condensed matter systems, such as lattices of trapped condensates, have enabled studies on non-linear effects in many-body systems where exchange of particles between lattice nodes is effectively instantaneous. Here, we demonstrate a regime of macroscopic matter-wave systems, in which ballistically expanding condensates of microcavity exciton-polaritons act as picosecond, microscale non-linear oscillators subject to time-delayed interaction. The ease of optical control and readout of polariton condensates enables us to explore the phase space of two interacting condensates up to macroscopic distances highlighting its potential in extended configurations. We demonstrate deterministic tuning of the coupled-condensate system between fixed point and limit cycle regimes, which is fully reproduced by time-delayed coupled equations of motion similar to the Lang-Kobayashi equation.
\end{abstract}

\maketitle

\section*{Introduction}

Time-delay is widespread in nature and occurs when the constituents of a given system interact via signals with a finite propagation time~\cite{mackey1977oscillation}. If the characteristic timescale of the system, such as the period of a simple pendulum, is much longer than the signal propagation time, then one arrives at familiar examples such as Huygens clock synchronisation described by instantaneous interactions~\cite{oliveira2015huygens}. When propagation times are appreciably long, the role of the system's history is enhanced and the interactions are said to be time-delayed. Such systems dictate the neurological function of our brains, affect traffic flow, influence economic activities, define population dynamics of biological species, regulate physiological systems, determine the stability of lasers, and have application in control engineering~\cite{erneux2009applied,atay2010complex}. Besides their ubiquity in nature and science, coupled systems with continuous time-delayed interactions exhibit interesting mathematical properties such as an infinite dimensional state space, i.e. for a fixed time-delay $\tau$ there are infinitely many initial conditions of the system in the time interval $-\tau \leq t \leq 0$ needed to predict the dynamics for $t>0$~\cite{schuster1989mutual}. Time-delayed interaction or self-feedback is known to greatly increase the dynamical complexity of a system, giving rise to chaotic motion, chimera states~\cite{Larger_PRL2013}, as well as being able to both stabilise and destabilise fixed point and periodic orbit solutions~\cite{atay2010complex}.

Technological applications of delay-coupled systems appear in diverse areas such as control engineering~\cite{scholl2008handbook}, high speed random-bit generation~\cite{uchida2008fast}, secure chaos communication~\cite{argyris2005chaos}, but have also recently emerged in machine learning, where the demand for neuro-inspired computing units has led to various hardware realisations of artificial neural networks~\cite{van2017advances}. The intrinsic high-dimensional state space of delay-coupled systems enables an efficient platform for tasks such as pattern recognition, speech recognition, and time series prediction as demonstrated in electronic~\cite{Appeltant2011}, photonic ~\cite{larger2012photonic,Brunner_NatCom2013,Vandoorne2014}, and optoelectronic~\cite{paquot2012optoelectronic,Martinenghi_PRL2012,Larger_PRX2017} systems.

In this work, we demonstrate the prospect of an ultrafast microscale platform for engineering time-delayed coupled oscillator networks based on condensates of microcavity exciton-polaritons (from here on  polaritons)~\cite{kavokin_microcavities_2007, li_tunable_AQT2016}. Polaritons are bosonic quasi-particles that can undergo a power-driven quantum phase transition to a macroscopically occupied state with long-range phase coherence~\cite{Kasprzak_Nature2006}. This phase transition is associated with the formation of a polariton condensate, a non-linear matter-wave quantum fluid, which differs from classical cold atom Bose-Einstein condensates, and liquid light droplets~\cite{wu_cubic-quintic_2013, zhang_particlelike_2019} due to its non-equilibrium nature. One of the greatest advantages of polariton condensates for optoelectronic applications is the easy implementation of arbitrary geometries, or graphs, using adaptive optical elements for the excitation laser beam ~\cite{Berloff2017a}. We show controllable tuning between steady state (single-colour) and limit-cycle (two-colour) regimes of the two coupled condensate system and explain our observations through time-delayed equations of motion.

\section*{Results}  
\subsection*{Time-delayed coupled matter-wave condensates}
\begin{figure}[!t]
	\center
	\includegraphics[]{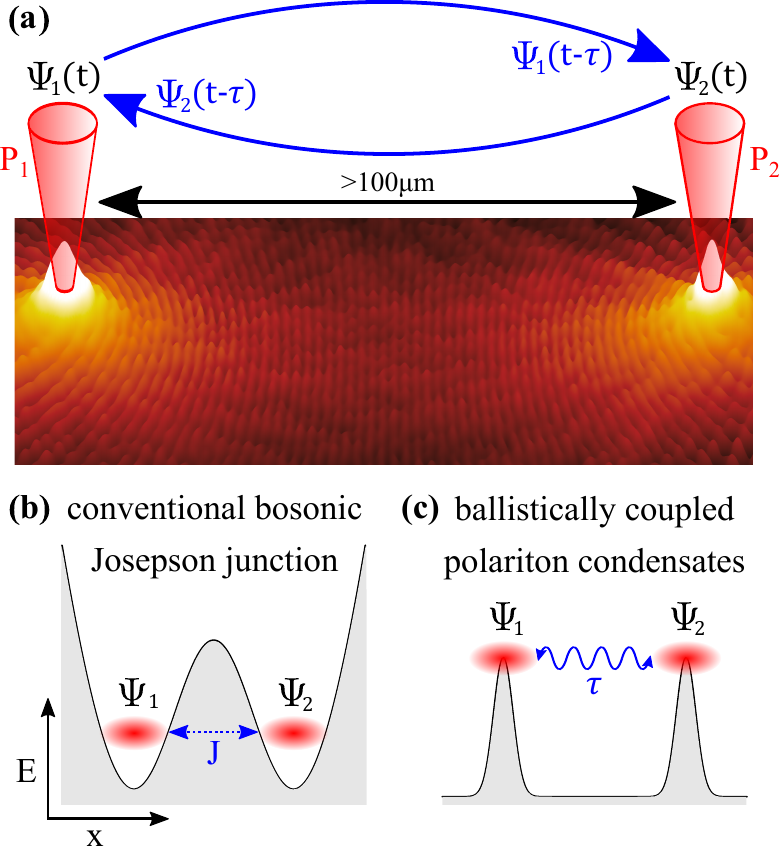}
	\caption{Macroscopically coupled matter-wave condensates. (a) Schematic showing the cavity plane with two focusing laser beams (red cones) which are spatially displaced by a distance $d\approx 114\;\mathrm{\upmu m}$ and the experimentally observed photoluminescence of two synchronised, ballistically expanding and interfering condensate centres. Finite particle transfer time $\tau$ results in time-delayed interaction between condensates $\Psi_{1}$ and $\Psi_{2}$. Unlike in the case of a conventional bosonic Josephson junction (b) the reported time-delayed coupling mechanism between two tightly-pumped polariton condensates (c) is not mediated by a tunneling current $J$ (blue dashed line) but by a radiative transfer of particles (blue wavy line).}
	\label{Fig1}
\end{figure}

Conventionally, networks and lattices of condensates consisting of cold atoms~\cite{bloch2005ultracold}, Cooper pairs~\cite{cataliotti2001josephson}, photons~\cite{dung2017variable} or polaritons~\cite{kim2011dynamical,Abbarchi_NatPhy2013} are studied in trap geometries, weakly coupled via tunneling currents, that allow for the study of solid state physics phenomena such as superfluid-Mott phase transitions~\cite{greiner2002quantum}, magnetic frustration~\cite{struck2011quantum}, $PT$-symmetric non-linear optics~\cite{Zhang_PRL2016}, and Josephson physics~\cite{albiez2005direct, Abbarchi_NatPhy2013}. Here we investigate the inverse case, wherein polariton condensates are freely expanding from small (point-like) sources  experiencing dynamics reminiscent to macroscopic systems such as time-delayed coupled semiconductor lasers ~\cite{Soriano_RevModPhy2013}. Ballistic expansion extends over two orders of magnitude beyond the pump beam waist, and occurs due to the repulsive potential formed by the uncondensed exciton reservoir injected by the non-resonant pump~\cite{Wouters_2008PRB}. In the case of spatially separated polariton condensates, propagation from one condensate centre to another results in a substantial phase accumulation, interpreted as a retardation of information flow between the condensates. To date, coupled polariton condensates were restricted to distances $d \lesssim 40\;\mathrm{\upmu m}$~\cite{christmann2014oscillatory, Ohadi2016,Berloff2017a}, wherein coupling was assumed to be instantaneous although propagation time was comparable or longer than the characteristic timescale of the systems. To unravel the role of time-delayed interactions, we demonstrate the synchronisation between two tightly-pumped condensates separated by up to $d \approx 114\;\mathrm{\upmu m}$, as shown in Fig.~\ref{Fig1}(a). Figures \ref{Fig1}(b) and (c) compare schematically the conventional regime of coupled condensates separated by a potential barrier and described by a tunneling current $J$, to the macroscopically coupled driven-dissipative matter-wave condensates interacting via radiative particle transfer subject to finite propagation time $\tau$.

\subsection*{Dynamics of two interacting condensates}
We utilise a strain compensated semiconductor microcavity to enable uninhibited ballistic expansion of polaritons over macroscopic distances~\cite{cilibrizzi2014polariton}. We inject a polariton dyad with identical Gaussian spatial profiles, with full-width-at-half-maximum (FWHM) of $\approx 2\;\mathrm{\upmu m}$, using non-resonant optical excitation at the first Bragg minima of the reflectivity stop-band, and employ spatial light modulation to continuously vary the separation distance from $6\;\mathrm{\upmu m}$ to $94\;\mathrm{\upmu m}$, while keeping constant the excitation density of each pump beam at $P_{1,2} \approx 1.5 \times P_{\mathrm{thr}}^{(1)}$, where $P_{\mathrm{thr}}^{(1)}$ is the condensation threshold power of an isolated condensate. For each separation distance $d$ (more than 400 positions), we record simultaneously the spatial profile of the photoluminescence (real-space), the dispersion (energy vs in-plane momentum in the direction of propagation), and the photoluminescence in reciprocal (Fourier) space (see Supplementary Movie 1). We observe that opposite to the two-fold hybridisation of two evanescently coupled condensates \cite{Abbarchi_NatPhy2013}, the macroscopically coupled system is characterised by a multitude of accessible modes of even and odd parity (i.e., $0$ and $\pi$ phase difference between condensate centres) that alternate continuously between opposite parity states with increasing dyad separation distance. For a range of separation distances only one resonant mode is present in the gain region of the dyad, wherein the polariton dyad is stationary, occupying a single energy level. Between the separation distances, wherein only one mode is present, we observe the coexistence of two resonant modes of opposite parity resulting in non-stationary periodic states.

\begin{figure}[!t]
	\center
	\includegraphics[]{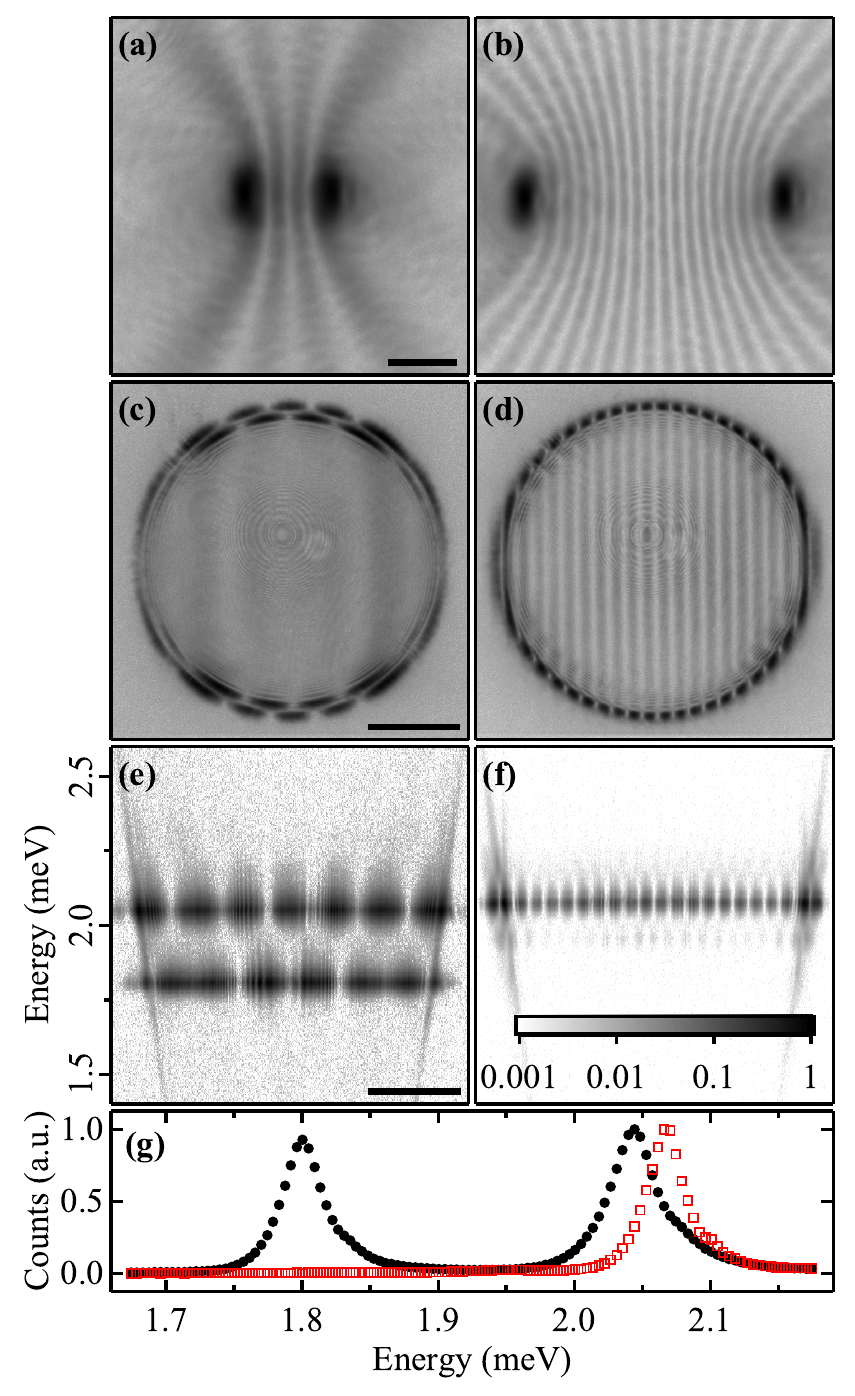}
	\caption{Emission characteristics of two non-resonantly pumped polariton condensates. Measured (a,b) real-space photoluminescence $|\Psi(x,y)|^2$, (c,d) momentum-space photoluminescence $|\Psi(k_{x},k_{y})|^2$ and (e,f) spectrally-resolved momentum-space photoluminescence along $k_y=0$ for two condensates with separation distances (a,c,e) $d=12.7\;\mathrm{\upmu m}$ and (b,d,f) $d=37.3\;\mathrm{\upmu m}$, respectively. (g) Normalised spectra, which are obtained by integrating (e) and (f), are illustrated with black dots and red hollow squares.  For better visibility of low-intensity features all images are illustrated in logarithmic grey-scale saturated below 0.001 as shown in (f). Scale bars in (a,c,e) correspond to ($10\;\mathrm{\upmu m}$,$1\;\mathrm{\upmu m}^{-1}$, $1\;\mathrm{\upmu m}^{-1}$).}
	\label{Fig2}
\end{figure}

In Fig.~\ref{Fig2} we present an example of the two different regimes, stationary and non-stationary states. Figures~\ref{Fig2}(a,b) show real-space photoluminescence, Fig.~\ref{Fig2}(c,d) show Fourier-space photoluminescence and Fig.~\ref{Fig2}(e,f) show the polariton dispersion along the axis of the dyad with separation distances of (a,c,e) $d=12.7\;\mathrm{\upmu m}$ and (b,d,f) $d=37.3\;\mathrm{\upmu m}$. Fig.~\ref{Fig2}(g) depicts the integrated spectra of Fig.~\ref{Fig2}(e,f) using black dots and red squares, respectively. Absolute values of the energy levels are given as a blueshift with respect to the ground-state of the lower polariton branch. Although from the clear fringe pattern in Fig.~\ref{Fig2}(a) one could infer that only one mode of even parity is present, it is through the simultaneous recording of either the Fourier-space or the dispersion that the coexistence of an odd parity mode becomes apparent. Each of the two modes has a well defined but opposite parity to the other. We note that as long as we maintain the mirror symmetry of the system by pumping both of the two condensate centres with the same power $P_1=P_2$, we do not observe formation of non-trivial phase configurations $\phi_{1}-\phi_2 \neq 0,\pi$.

\begin{figure}[!t]
	\center
	\includegraphics[]{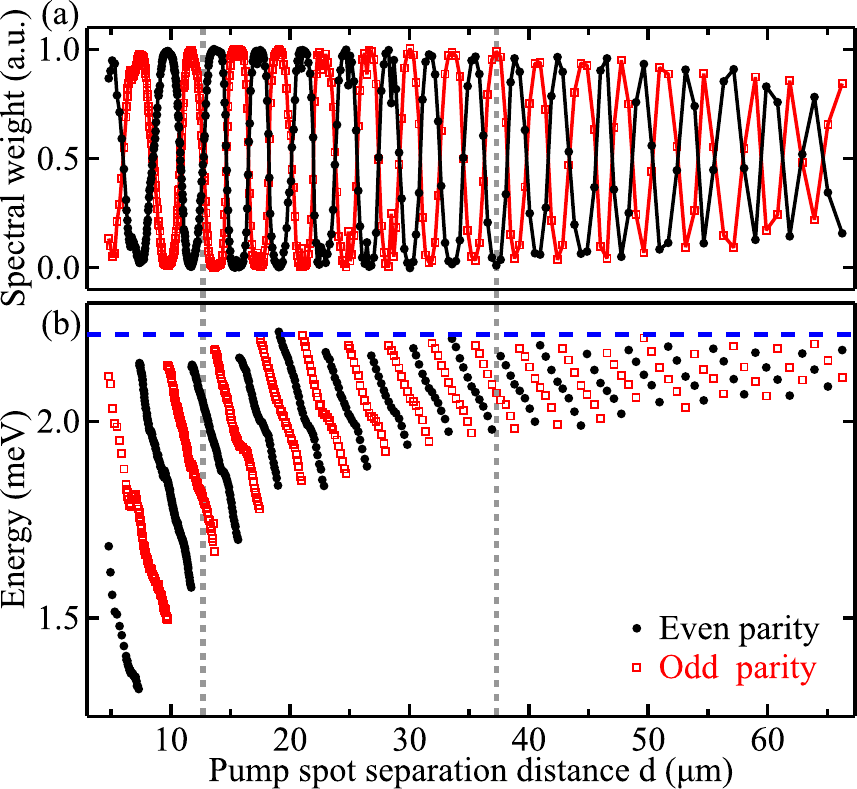}
	\caption{ Spectra of ballistically coupled polariton condensates. (a) Spectral weight and (b) measured blueshift of even (black dots) and odd (red hollow squares) parity states formed by the coupling of two spatially separated polariton condensates with distance $d$. The energy level of a single condensate pumped with the same non-resonant excitation power density, i.e. $1.5$ times the condensation threshold of a single condensate, is illustrated with a horizontal blue dashed line. The vertical dashed lines at $d=12.7\;\mathrm{\upmu m}$ and $d=37.3\;\mathrm{\upmu m}$ indicate a two-colour and single-colour state, respectively.}
	\label{Fig3}
\end{figure}

To unravel the dynamics of the coupling on the separation distance between the two condensates, we obtain the spectral position, parity, and spectral weight of both energy-levels for pump spot separation distances from $d=5\;\mathrm{\upmu m}$ to $d=66\;\mathrm{\upmu m}$. We have recorded configurations with more than two occupied energy levels for several distances $d$, but with the relative spectral weight of the third peak less than a few percent. In the following, we focus our analysis to the two brightest energy levels for each configuration. The measured normalised spectral weights and spectral positions of the two states versus pump spot separation distance $d$ are illustrated in Fig.~\ref{Fig3} (a) and (b) using black dots for even parity states and red hollow squares for odd parity states, respectively. Figure~\ref{Fig3}(b) shows that the system follows an oscillatory behaviour in the spectral weights of the two parity states giving rise to continuous transitions between even and odd parity states interleaved by configurations exhibiting both parity states. Interestingly, each period of these oscillations in relative intensity (starting and ending with a vanishing spectral weight) displays an `energy branch' featuring a notable reduction in energy with increasing pump spot separation distance $d$ as is shown in Fig.~\ref{Fig3}(b). The energy of an isolated condensate is measured in the same sample space and its value $\approx 2.22\; \mathrm{meV}$, above the ground state energy of the lower polariton dispersion, is illustrated with a blue dashed horizontal line. We observe that the phase-coupling of two spatially separated condensate centres is dominated by a spectral red-shift with respect to the energy of an isolated condensate. The spectral size of each `energy branch', i.e., the measurable red-shift, is decaying from branch-to-branch with increasing pump spot separation distance $d$. The two dyad configurations exhibiting single-colour and two-colour states shown in Fig.~\ref{Fig2} are indicated with gray vertical dashed lines in Fig.~\ref{Fig3}.
Interestingly, it has been shown that the observation of phase-flip transitions, accompanied with changes in oscillation frequency to another mode are a universal characteristic of time-delayed coupled non-linear systems~\cite{schuster1989mutual, prasad2008universal}. Such dynamics can be associated with neuronal systems~\cite{adhikari2011time, barardi2014phase, dotson2016experimental}, and coupled semiconductor lasers~\cite{Soriano_RevModPhy2013,clerkin2014multistabilities}, but have also been demonstrated experimentally for other types of time-delayed coupled systems such as non-linear electronic circuits~\cite{reddy2000experimental,prasad2008universal}, living organisms~\cite{takamatsu2000time}, chemical oscillators~\cite{cruz2010phase} and candle-flame oscillators~\cite{manoj2018experimental}.

\begin{figure}[!t]
	\center
	\includegraphics[]{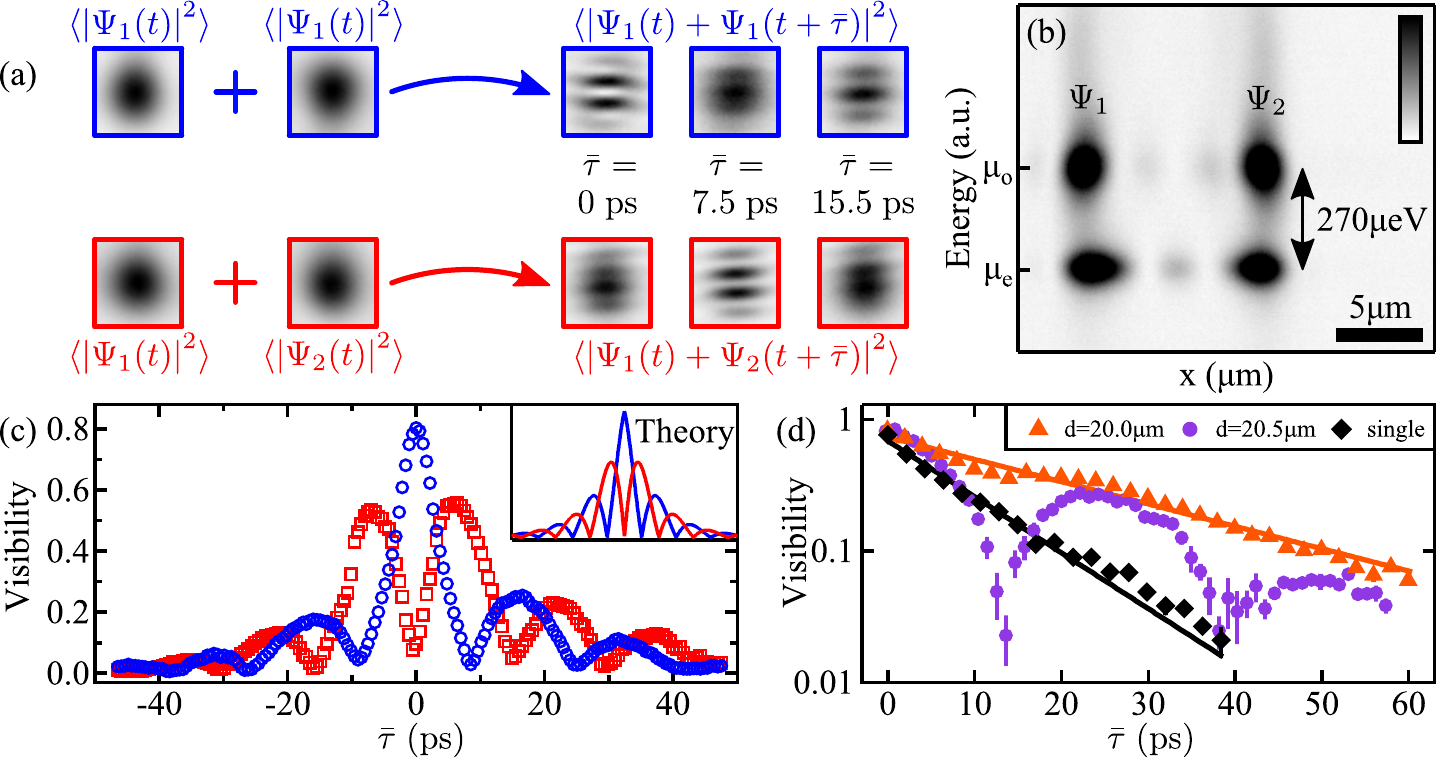}
	\caption{Periodic population transfer between two ballistically coupled polariton condensates. The emission of a polariton dyad with condensation centres $\Psi_{1,2}$ pumped with $P_{1,2}=1.5P^{(1)}_{\mathrm{thr}}$ at locations $(x=\pm d/2,y=0)$ with $d\approx 10.3 \; \upmu \mathrm{m}$ is interferometrically and spectrally investigated in (a-c). (a) Time integrated real space photoluminescence of both condensates and the interference patterns of one condensate centre $\Psi_1(t)$ interfering with a delayed version of itself $\Psi_1(t+\bar{\tau})$ or with a delayed version of the spatially displaced condensate $\Psi_2(t+\bar{\tau})$. (b) Corresponding spectrally resolved real-space photoluminescence along the axis of the dyad $y=0$. The colour-scale of the normalised counts is saturated above $0.5$ for better visibility. (c) Extracted fringe visibilities for the interference of same condensate centres (blue hollow circles) and opposite condensate centres (red hollow squares) versus relative time-delay $\bar{\tau}$ of the two interferometer arms. The inset in (c) illustrates the expected visibilities [Eq.~\eqref{time_average_config1} and~\eqref{time_average_config2}] multiplied with an exponentially decaying envelope. (d) Comparison of the temporal decay of coherence extracted from the interference of the same condensate $\langle \left| \Psi_1(t)+\Psi_1(t+\bar{\tau}) \right|^2 \rangle$ for two coupled condensates with separation distances $d=20\;\upmu\mathrm{m}$ and $d=20.5\;\upmu\mathrm{m}$, as well as a single isolated condensate. In all three cases each condensate is pumped equally with $P\approx 1.7 P^{(1)}_\mathrm{thr}$. The error bars are calculated as the standard deviation of the extracted visibility within the full width at half maximum ($\approx 2\;\mathrm{\upmu m}$) of each condensate. Lines represent exponential fits for the decay of coherence for the single-mode polariton dyad and the single condensate, respectively.}
	\label{Fig4}
\end{figure}

Time averaged measurements over $\sim1000$ realisations of the system as presented in Fig.~\ref{Fig2}(a,c,e) cannot reveal whether the system of two coupled polariton condensates exhibits periodic dynamics in the form of two-colour states or whether it stochastically picks one of the two parity states. However, it is well known that the power spectrum of a time-dependent signal is related to its auto-correlation function by means of Fourier-transform (Wiener-Khinchin theorem). A two-colour solution, as shown in Fig.~\ref{Fig2}(g), results in the periodic disappearance and revival of the first order temporal coherence function $g^{(1)}(\bar{\tau})$. To characterise the temporal evolution of $g^{(1)}(\bar{\tau})$, we perform interferometric measurements of the two coupled condensates and extract the fringe visibility $\mathcal{V}$ which - up to a normalisation factor - is proportional to the magnitude of the coherence function $\left| g^{(1)} \right|$.  Assuming two coexisting states of opposite parity but equal spectral weight, one may simplify the complex amplitudes of the two coupled condensates $\Psi_1$ and $\Psi_2$ as
\begin{align}
\Psi_1(t) &= \psi_0 \left( e^{-i\mu_{\mathrm{e}}t/\hbar} + e^{-i\mu_{\mathrm{o}}t/\hbar}e^{i\phi)} \right), \label{timedependent_condensate_1} \\
\Psi_2(t) &= \psi_0 \left( e^{-i\mu_{\mathrm{e}}t/\hbar} - e^{-i\mu_{\mathrm{o}}t/\hbar}e^{i\phi)} \right), \label{timedependent_condensate_2}
\end{align}
with time-independent complex amplitude $\psi_0$ and relative phase $\phi$ between the two modes of opposite parity and energy $\mu_{\mathrm{e}} ,\mu_{\mathrm{o}}$ for the even and odd mode respectively. The mixture of two modes, with energy splitting $\hbar\Delta = \left| \mu_{\mathrm{e}} - \mu_{\mathrm{o}} \right|$, causes an antiphase temporal beating in the intensities $\left| \Psi_{1,2} \right|^2$ similar to oscillatory population transfer in coupled bosonic Josephson junctions. Therefore, while the occupation amplitude of both condensate centres oscillates with a period $T = 2\pi/\Delta$, they feature a relative phase shift of $\pi$ due to mixture of even and odd parity states. We use a Michelson interferometer, comprising of a retroreflector mounted on a translational stage and measure both the time averaged interference of the photoluminescence of the same condensate $I_{11}(\bar{\tau}) = \left< \left|  \Psi_1(t) + \Psi_1(t+\bar{\tau}) \right| ^2 \right>$ as well as the time averaged interference of the photoluminescence of opposite condensates $I_{12}(\bar{\tau}) = \left< \left| \Psi_1(t) + \Psi_2(t+\bar{\tau}) \right| ^2 \right>$, where $\bar{\tau}$ is the relative time-delay controlled by the adjustable position of the retroreflector. Examples of the interferometric images are illustrated in Fig.~\ref{Fig4}(a) for two condensates with a separation distance of $d=10.3\;\mathrm{\upmu m}$, which demonstrates condensation in both an even and an odd parity state as shown in Fig.~\ref{Fig4}(b). Assuming the spectral composition as noted in Eqs.~\eqref{timedependent_condensate_1} and~\eqref{timedependent_condensate_2} the corresponding interferometric visibilities $\mathcal{V}$ can be written as
\begin{align}
\mathcal{V}_{11} = \left| \cos(\Delta \bar{\tau}/2) \right| , \label{time_average_config1} \\
\mathcal{V}_{12} = \left| \sin(\Delta \bar{\tau}/2) \right| .  \label{time_average_config2}
\end{align}
The experimentally extracted visibilities $\mathcal{V}_{11}(\bar{\tau})$ and $\mathcal{V}_{12}(\bar{\tau})$ versus relative time-delay $\bar{\tau}$ are illustrated in Fig.~\ref{Fig4}(c) with blue circles and red squares, respectively. In agreement with the predicted correlations [Eqs.~\eqref{time_average_config1} and~\eqref{time_average_config2}] we observe high fringe visibility $\mathcal{V}_{11}(0) > 0.8 $ for the interference of one condensate with itself and almost vanishing fringe visibility $\mathcal{V}_{12}(0) < 0.1 $ for the interference of opposite condensates at zero time-delay. Furthermore, we find periodic disappearance and revival of fringe visibility with period $T\approx 15.3\;\mathrm{ps}$, which is consistent with the observed energy-splitting of the two modes $\hbar \Delta = 270\;\mathrm{\upmu eV}$ [see Fig.~\ref{Fig4}(b)]. The relative phase shift between the two visibilities $\mathcal{V}_{11}(\bar{\tau})$ and $\mathcal{V}_{12}(\bar{\tau})$ is a direct result of the periodic population transfer between the two condensates. For comparison the inset in Fig.~\ref{Fig4}(c) shows the expected visibilities [Eqs.~\eqref{time_average_config1} and~\eqref{time_average_config2}] multiplied with an exponential decay accounting for the finite coherence time of the system.

In Fig.~\ref{Fig4}(d) we show the decay of temporal coherence for two coupled condensates in a single-colour state ($d=20\;\mathrm{\upmu m}$), a two-colour state ($d=20.5\;\mathrm{\upmu m}$) and for an isolated condensate excited with the same pump power density. We note that the coherent exchange of particles in both regimes of the coupled condensate system results in an enhanced coherence time. For the single-colour state ($d=20\;\mathrm{\upmu m}$) and the isolated condensate the coherence time $\bar{\tau}_{\mathrm{c}}$ can be extracted from exponential fits yielding $25.5\;\mathrm{ps}$ and $10.2\;\mathrm{ps}$, respectively. 

\subsection*{Numerical analysis}

The dynamics of polariton condensates can be modelled via the mean field theory approach where the condensate order parameter $\Psi(\mathbf{r},t)$ is described by a 2D semiclassical wave equation often referred as the generalised Gross-Pitaevskii equation coupled with an excitonic reservoir which feeds non-condensed particles to the condensate~\cite{Wouters_PRL2007}. In Supplementary Note 1 we give numerical results of the spatiotemporal dynamics of the dyad reproducing semi-quantitatively the experimentally observed results. We also provide a supplemental animation showing the calculated evolution of the two-colour condensate (Supplementary Movie 2).

In the following, we show that our experimental observations are described as a system of time-delayed coupled non-linear oscillators. For simplicity we consider the 1D Schrödinger equation corresponding to the problem of $s$-wave scattering of the condensate wavefunctions with $\mathrm{\delta}$-shaped complex-valued, pump induced, potentials. We start by characterising the energies of the time-independent non-hermitian single particle problem:
\begin{equation} \label{eq.DeltaPotentialToyModel}
E \Psi(x) = \left( -\frac{\hbar^2\partial_x^2}{2m} + V(x)  - i \frac{\hbar \gamma_{\mathrm{c}}}{2} \right) \Psi(x),
\end{equation}
where $V(x)$ describes complex-valued $\delta$-shaped potentials separated by a distance $d$, $V(x)= V_0  \left(\delta(x+d/2)+\delta(x-d/2)\right)$, and $V_0$ lies in the first quadrant of the complex plane, i.e. repulsive interactions and gain. The eigenfunctions of Eq.~\eqref{eq.DeltaPotentialToyModel} describing normalisable solutions of outwards propagating waves from the potential (non-resonant pump) centres are written as,
\begin{equation} \label{Eq.1DToyModel_WavefunctionSolution}
	\Psi(x) = \begin{cases}
	A e^{-ikx}, & x \leq -d/2 \\
	B e^{ikx} + C e^{-ikx}, & |x|<d/2 \\
	D e^{ikx},  & x \geq d/2
	\end{cases}
\end{equation}
where $k$ also belongs to the first quadrant of the complex plane. This problem is well known for the case of lossless attractors ($\Re{(V_0)}<0$, $\Im{(V_0)}=0$) describing electron states in a 1D diatomic Hydrogen molecule ion. The resonance condition of the system is written,
\begin{equation} \label{Eq.ResonanceCondition}
\left[ \frac{V_0}{\frac{i\hbar ^2 k}{m} - V_0} e^{ikd} \right]^2 = 1.
\end{equation}
The solutions of Eq.~\eqref{Eq.ResonanceCondition} can explicitly be written as,
\begin{figure}[!t]
	\center
	\includegraphics[width=8.6cm]{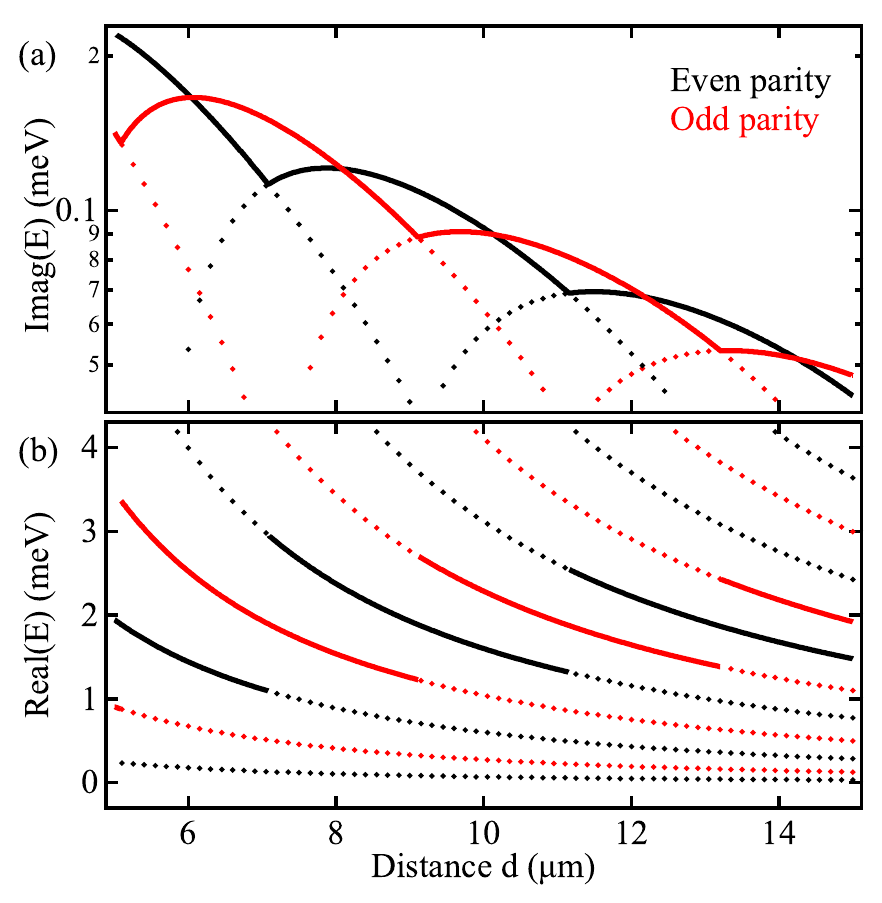}
	\caption{ Resonances of two complex-valued $\delta$-shaped potentials. Calculated (a) imaginary part and (b) real part of the eigenvalues $E$ in Eq.~\eqref{eq.dispersion} for $\Re{(V_0)} = 1\; \mathrm{meV}\; \mathrm{\upmu m}$, $ \Im{(V_0)} = 2\mathrm{meV}\; \mathrm{\upmu m}$, $m=0.28\;\mathrm{meV}\; \mathrm{ps}^2 \; \mathrm{\upmu m}^{-2}$, $\gamma_{\mathrm{c}}=1/5.5\;\mathrm{ps}^{-1}$. The two solutions with largest gain (imaginary part of $E$) are illustrated as lines, other solutions are shown as dotted branches.}
	\label{Fig5}
\end{figure}
\begin{equation} \label{eq.TrEq}
k_{n,\pm} = -i \tilde{V} + \frac{i}{d} W_n(\mp d \tilde{V} e^{d \tilde{V}}), \quad n \in \mathbb{Z},
\end{equation}
with $\tilde{V} = mV_0/\hbar^2$. Equation~\eqref{eq.TrEq} describes infinitely many solutions of the system of even $(+)$ and odd $(-)$ parity, where $W_n$ are the branches of the Lambert $W$ function. The corresponding complex-valued eigenvalues,
\begin{equation} \label{eq.dispersion}
E_{n,\pm}=\frac{\hbar^2 k_{n,\pm}^2}{2m} - \frac{i\hbar\gamma_{\mathrm{c}}}{2},
\end{equation}
are illustrated in Fig.~\ref{Fig5} and qualitatively reproduce the experimental findings of the multiple energy branches shown in Fig.~\ref{Fig3}. We interpret the experimental occurrence of predominantly two lasing modes with the behaviour of the imaginary values of $E_n$ for the simplistic 1D-toy model. While there are periodically alternating regions of even and odd-parity solutions dominating the gain, we expect distances at which two modes (of opposite parity) have equal gain and thus can operate with equal intensity. It is worth noting that the Lambert $W$ function naturally arises for problems involving delay differential equations~\cite{Asl2003}.

The time-dependent problem can be formulated as a superposition of two displaced, normalisable, and ballistically propagating waves $\psi_{1,2}(x)$, each emerging from one of the condensate centres,
\begin{equation} \label{Eq.Superposition}
\Psi(x,t) = c_1(t)  \psi_{1}(x) + c_2(t) \psi_{2}(x).
\end{equation}
Here, in analogy with Eq.~\eqref{Eq.1DToyModel_WavefunctionSolution}, the normalised ansatz is written as $\psi_{1,2}(x) = \sqrt{\kappa} e^{ik|x \pm d/2|}$, with a complex-valued wavevector $k=k_{\mathrm{c}} + i \kappa$. When the coupling between condensates is weak, i.e. small $\xi = \exp(- \kappa d)$, one can omit all terms of order $\mathcal{O}(\xi^2)$ and higher. Then plugging Eq.~\eqref{Eq.Superposition} into the time-dependent form of Eq.~\eqref{eq.DeltaPotentialToyModel} and integrating out the spatial degrees of freedom, assuming that Eq.~\eqref{Eq.ResonanceCondition} is satisfied, one gets (see Supplementary Note 2),
\begin{equation} \label{eq.CoupledCondensates_2}
i\hbar \dot{c}_i  =  \left[ \frac{\hbar^2 k^2}{2m} - i\frac{\hbar\gamma_{\mathrm{c}}}{2}  + \kappa \left(V_0 -  \frac{i\hbar ^2 k}{m} \right) \right] c_i  +   V_0 \kappa e^{ikd} c_j.
\end{equation}
where $j=3-i$ and $i=1,2$ are the condensate indices. When setting $c_{i} = \pm c_{j}$, and solving for stationary states, the above equation recovers the exact resonant solutions dictated by Eq.~\eqref{eq.TrEq}.  Equation~\eqref{eq.CoupledCondensates_2} then shows that inter-condensate interaction is in the form of a coherent influx of particles from condensate $j$ onto the condensate centre of condensate $i$ (and vice versa), with a phase retardation of $k_{\mathrm{c}}d$. When $c_i$ and $c_j$ oscillate at a fixed frequency $\omega$ we can transform the phase-shifting term $\exp(ik_{\mathrm{c}}d)$ into an effective time-delay,
\begin{equation} \label{Eq.TimeDelay}
e^{ik d} c_j(t) = e^{- \kappa d} c_j(t-\tau).
\end{equation}
The time-delay $\tau$ corresponds to an interaction lag between condensate centres $i$ and $j$ caused by their spatial separation and is given by $\tau = k_{\mathrm{c}}d/\omega$. In the case of weak coupling, i.e. small changes in oscillation frequency $\omega$ compared to the frequency $\omega_0$ of a single unperturbed condensate, the time-delay is approximately proportional to the dyad separation distance $d$, i.e.  $\tau \approx k_{\mathrm{c},0}d/\omega_0$ where we also use the notation $k_0 = k_{\mathrm{c},0} + i \kappa_0$ for the wavevector of the single unperturbed condensate. The exponentially decaying term on the right-hand side of Eq.~\eqref{Eq.TimeDelay} accounts for the one-dimensional spatial decay of particles propagating in between the two condensate centres. Introducing local non-linear interactions, reservoir gain and blueshift one can write the full non-linear equation of motion for the two coupled condensates as~\cite{Wouters_PRL2007},
\begin{align}
i \dot{c}_i &=  \left[ \Omega  + \left(g + i \frac{R}{2}\right) n_i + \alpha  |c_i|^2  \right] c_i  +  J e^{i\beta } c_j(t-d/v),  \label{eq.CoupledCondensates_condensates}  \\
\dot{n}_i & = -(\Gamma_{\mathrm{A}} + R |c_i|^2) n_i + P.   \label{eq.CoupledCondensates_reservoir}
\end{align}
Here, $n_i$ correspond to the pump induced exciton reservoirs providing blushift and gain into their respective condensates, $g$ is the polariton-reservoir interaction strength, $R$ is the rate of stimulated scattering of polaritons into the condensate from the active reservoir, $\alpha$ is the interaction strength of two polaritons in the condensate, $v = \omega_0/k_{\mathrm{c},0}$ is the phase velocity of the polariton wavefunction, and $\Gamma_{\mathrm{A}}$ is the radiative decay rate of the bottleneck reservoir excitons. The parameter $\Omega = \Omega_0 - i \Gamma$ captures the self energy of each condensate which will, in general, have a contribution from a background of optically inactive dark excitons generated at the pump spot, and $\Gamma$ denotes the effective linewidth of the polaritons expanding away from the pump spot. The complex valued coupling is written $J \exp{(i \beta ) } =  V_0 \kappa \exp{(-\kappa d)}$ for brevity. Equation~\eqref{eq.CoupledCondensates_condensates} is then in the form of a discretised complex Gross-Pitaevskii equation but with time-delayed interaction between the bosonic particle ensembles; which greatly increases the dimensionality of phase-space and complexity of the coupled system. Our system then has strong similarity with the famous Lang-Kobayashi equation~\cite{Kobayashi_JQE1980, kozyreff2000global} where in our case each condensate acts as a radiating antenna of symmetrically expanding waves. The two spatially separated antennas interfere and maximise their gain by adjusting both their common frequency and their relative phase difference $\phi = \arg{(c_i^* c_j)}$. Similarities of the dynamics of coupled polariton condensates to equations of motion in the form of the Lang-Kobayashi equation was discussed in theoretical works recently for instantaneously coupled condensates with complex-valued couplings~\cite{kalinin2019polaritonic}. Unlike trapped ground state bosonic systems, such as cold atoms, the ballistically expanding polariton matter-wave condensates necessarily experience time-delayed coupling since $k_{\mathrm{c},0} d \gg 1$ similar to inter-cavity coupling of semiconductor lasers~\cite{Soriano_RevModPhy2013}.

\begin{figure}[!t]
	\center
	\includegraphics[]{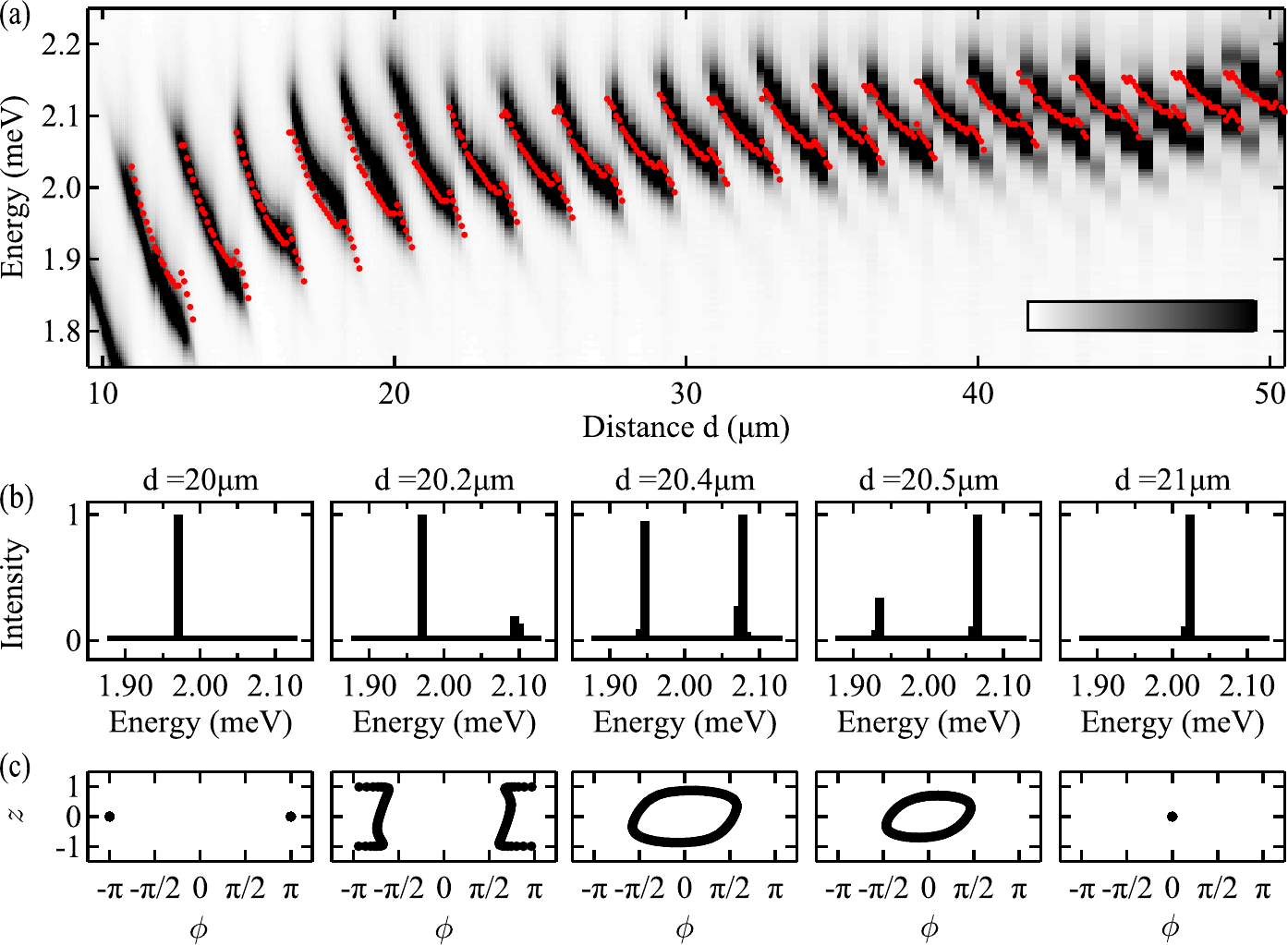}
	\caption{Periodic parity-transitions described by time-delayed coupled oscillators. (a) Comparison of experimentally measured spectra of two coupled polariton condensates coupled with separation distance $d$ and numerically simulated spectral peaks using Eqs.~\eqref{eq.CoupledCondensates_condensates},~\eqref{eq.CoupledCondensates_reservoir} and \eqref{eq.CouplingParameters}. Experimentally measured spectra are normalised for each distance $d$ and the grey-scale colour-map is saturated above $0.5$ for better visibility. (b) and (c) illustrate the spectral decomposition and phase-diagram for the anti-phase to in-phase transition of the system by increasing the distance $d$ from $20\;\mathrm{\upmu m}$ to $21\;\mathrm{\upmu m}$. Simulation parameters: $\hbar \Omega = (1.22-i 0.5)\; \mathrm{meV}$, $\hbar \alpha = 0.1\;\mathrm{\upmu eV}$, $\hbar R= 0.5\;\mathrm{\upmu eV}$, $\hbar g= 0.5\;\mathrm{\upmu eV}$, $v = 1.9\;\mathrm{\upmu m}\;\mathrm{ps}^{-1}$, $P=100\;\mathrm{ps}^{-1}$, $\Gamma_{\mathrm{A}}=0.05\;\mathrm{ps}^{-1}$, $\hbar J_0 = 1.1\;\mathrm{meV}$, $k_0=(1.7+i 0.014) \mathrm{\upmu m}^{-1}$ and $\beta = -1$.}
	\label{Fig.5}
\end{figure}

In order to accurately reproduce experimentally observed spectra using Eqs.~\eqref{eq.CoupledCondensates_condensates} and~\eqref{eq.CoupledCondensates_reservoir} we fit the distance dependence of the coupling amplitude $J(d)$ to the spatial envelope of a single condensate. It is known that in a 2D system the $0$-order Hankel function of the first kind describes the cylindrically symmetric radial outflow of particles in the linear regime~\cite{Wouters_2008PRB}, therefore we choose
\begin{equation} \label{eq.CouplingParameters}
J(d)  = J_0 \left| H^{(1)}_0(k_0d) \right|,
\end{equation}
where $J_0$ is a parameter describing the coupling strength (see Supplementary Note 3). Numerical integration of Eqs.~\eqref{eq.CoupledCondensates_condensates} and~\eqref{eq.CoupledCondensates_reservoir} is computed for separation distances $d>10\;\mathrm{\upmu m}$ using Gaussian white noise as initial conditions which, in close analogy to the experimental findings, reproduces periodic parity-flip transitions accompanied by cyclic solutions in the transition region. Fig.~\ref{Fig.5}(a) depicts the experimentally measured and normalised emission spectra of the polariton dyad versus pump spot separation distance $d$ on a grey-scale colour-map, for which the extracted two most dominant spectral peaks are depicted in Fig.~\ref{Fig3}. The red dots represent the numerically calculated spectral peak from Eqs.~\eqref{eq.CoupledCondensates_condensates} when in single mode operation, or the two most dominant spectral peaks when multiple spectral components exist; showing excellent agreement with experiment. As an example, we illustrate a continuous transition of the system from an anti-phase to in-phase state described by the time-delayed coupled model in Fig.~\ref{Fig.5}(b) and (c) showing the corresponding spectral decomposition and phase-space diagrams of the system for an increasing set of distances $d$ from $20\;\mathrm{\upmu m}$ to $21\;\mathrm{\upmu m}$. Similar stationary and oscillatory behaviour, for separation distances $d=20\;\mathrm{\upmu m}$ and $d=20.5\;\mathrm{\upmu m}$ respectively, is shown in Fig.~\ref{Fig4}(d). The numerically simulated phase-space diagrams depict periodic orbits in the transition region, involving periodic oscillations of the phase difference $\phi = \arg (c_1^* c_2)$ and population imbalance $z = (|c_1|^2-|c_2|^2)/(|c_1|^2+|c_2|^2)$, which is also confirmed by 2D-simulations of the generalised Gross-Pitaevskii equation (see Supplementary Note 1). We have verified through numerics that time-delay physics are indeed an accurate representation of the coupled condensate dynamics (see Supplementary Note 2).

We note that in the limit of fast active reservoir relaxation, $\Gamma_{\mathrm{A}}^{-1} \ll \Gamma^{-1}$, one can adiabatically eliminate Eq.~\eqref{eq.CoupledCondensates_reservoir} and introduce an effective non-linear term to Eq.~\eqref{eq.CoupledCondensates_condensates}, $(\alpha_\text{eff} - i\sigma)|c_i|^2$, accounting for polariton-polariton and polariton-reservoir interactions, as well as condensate gain saturation. The dynamical equations of two coupled condensates are then described by time-delayed coupled Stuart-Landau oscillators,
\begin{equation}\label{eq.CoupledCondensates_4}
i \dot{c}_i =  \left( \Omega  +  (\alpha_\text{eff} - i \sigma) |c_i|^2  \right) c_i  +  J e^{i\beta } c_j(t-d/v).
\end{equation}
Numerical analysis of Eq.~\eqref{eq.CoupledCondensates_4} is given in Supplementary Note 4 showing qualitative agreement with experiment.

\section*{Discussion}
We present an extensive experimental and theoretical study of a system of two ballistically expanding (untrapped) interacting polariton condensates, the fundamental building block of polariton graphs with higher connectivity. We demonstrate a regime for coupled matter-wave condensates, wherein the coupling is not instantaneous but mediated by a particle flow inherently connected with time retardation effects. We observe deterministic selection of steady state (single-colour) or dynamical (two-colour) modes of the system by controlling the separation distance between the condensates. Time-delay polaritonics potentially offers an ultrafast platform for simulating the dynamics of real-world systems of time-delayed coupled non-linear oscillators that appear in photonics, electronics and neural circuits. Given the high non-linearities of polaritons and the ease of optically imprinting multiple condensates of arbitrary geometries on planar microcavities, the system offers promising applications for neuromorphic devices based on lattices of history dependent, non-trapped, strongly interacting, polariton condensates with a wide range of coupling strengths, fast optical operations (input), dynamics (processing), and readout.

\section*{Methods}
\subsection*{Microcavity sample and experimental methods}
The microcavity sample used is a strain compensated $2 \lambda$ GaAs microcavity with three pairs of 6 nm InGaAs quantum wells embedded at the anti-nodes of the electric field~\cite{cilibrizzi2014polariton}. The intracavity layer contains a wedge and the position on the sample is chosen such that the cavity photonic mode is red-shifted from the excitonic mode at zero inplane wavevector ($|\mathbf{k}|=0$) by $\approx -5.5$ meV. For all experiments the microcavity is held in a cold finger cryostat (temperature $T\approx 6\;K$) and is optically pumped with a circularly polarised continuous wave monomode laser blue-detuned above the cavity stopband ($\lambda \approx 785\; \mathrm{nm}$). To prevent heating of the sample an acousto-optic modulator is used to generate square wave packets at a frequency of 10 kHz and duty cycle of $5\%$. A liquid crystal spatial light modulator (SLM) imprints a phase pattern such that when the beam is focused through the 0.4 numerical aperture microscope objective lens it excites the sample with the desired spatial geometry. The phase patterns are carefully designed so that when changing the pump separation distance $d$ of the two-spot excitation pattern, the diffraction efficiency of the SLM remains constant and both pump spots retain equal excitation power and width. The photoluminescence is collected in reflection geometry and spectrally resolved using an 1800 grooves/mm grating in a $750\;\mathrm{mm}$ spectrometer, which is equipped with a charge-coupled device (CCD). Real-space,  Fourier-space and dispersion images are acquired using exposure times in the order of milliseconds.
\subsection*{Interferometry}
A modified Michelson-interferometer, where one mirror is replaced with a retroreflector mounted on a translational stage, is used for measuring the interference and temporal coherence of the emission. By tilting the angle of the emission entering the interferometer we select to spatially overlap the emission of either opposing condensates or the same condensates onto a CCD camera. Interference fringe visibilities are extracted from the normalised $1^{\mathrm{st}}$ diffraction order of the computed discrete Fourier transform of each interference image.
\subsection*{Image processing}
Image displayed in Figure 1(a) has been digitally processed with a low-pass filter to increase the visibility of interference fringes.


\section*{Acknowledgements}
The authors are grateful to N. G. Berloff for fruitful discussions and acknowledge the support of the UK’s Engineering and Physical Sciences Research Council (grant EP/M025330/1 on Hybrid Polaritonics).

\section*{Author contributions}
P.G.L. led the research project. P.G.L., J.D.T and L.P designed the experiment. J.D.T. and L.P. carried out the experiments and analysed the data. J.D.T and H.S. developed the theoretical modelling. H.S performed numerical simulations. All authors contributed to the writing of the manuscript.

\section*{Data availability}
The data that support the findings of this study are openly available from the University of Southampton repository (DOI: 10.5258/SOTON/D1149)~\cite{dataset}.

\section*{Competing interests}
The authors declare no competing interests.

\newpage
\begin{center}
 \textbf{\Large Supplemental Information}
 \end{center}

\setcounter{equation}{0}
\setcounter{figure}{0}
\renewcommand{\theequation}{S\arabic{equation}}
\renewcommand{\thefigure}{S\arabic{figure}}
\renewcommand{\thesection}{S\arabic{section}}

\begin{figure}[!ht]
	\center
	\includegraphics[width=0.6\linewidth]{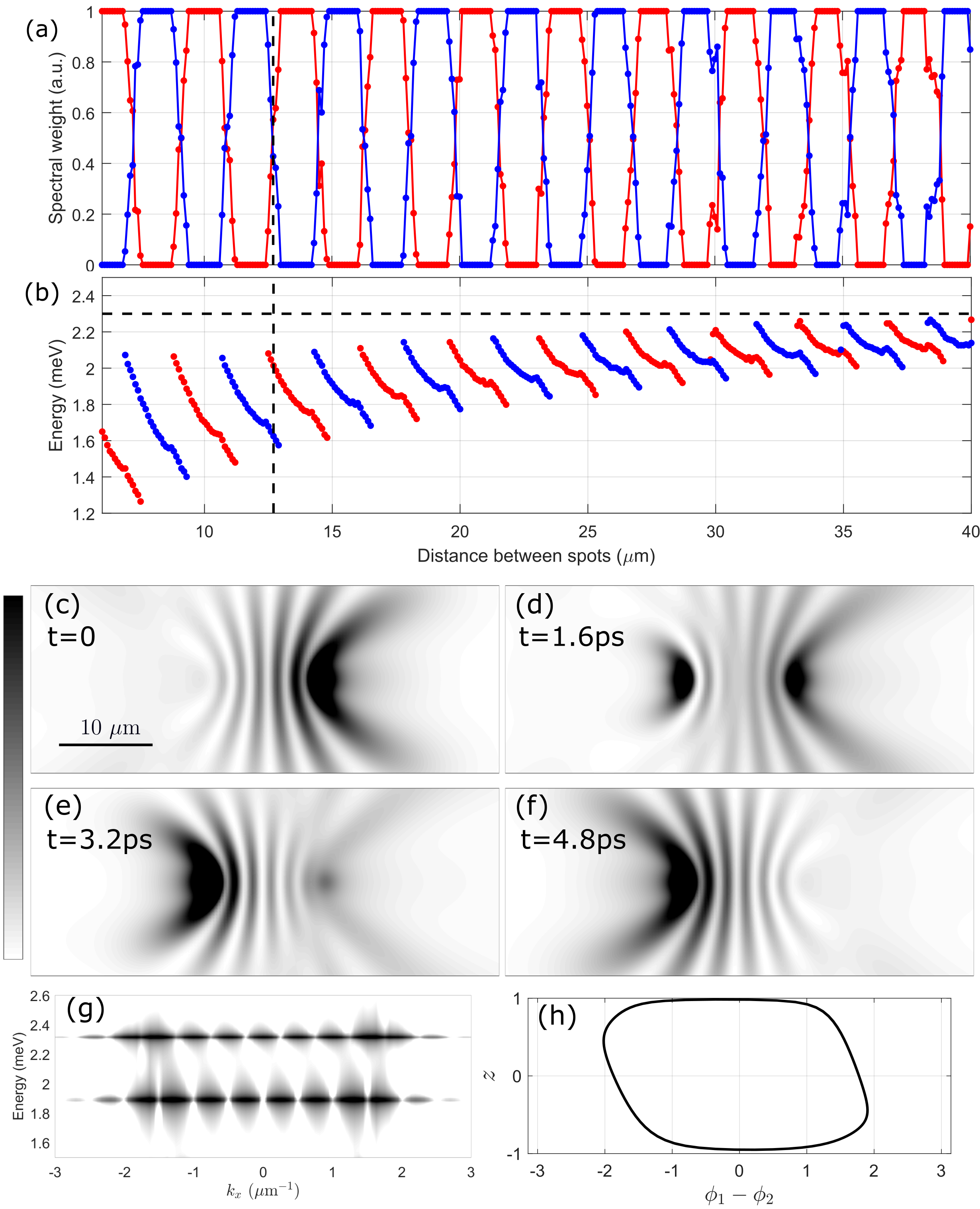}
	\caption{(a-b) Calculated relative visibility and spectral position of even (red) and odd (blue) parity states formed by the coupling of two spatially separated polariton condensates for $P_0 = 1.1P_\text{thr}^{(1)}$. The measured energy level of a single condensate pumped with the same non-resonant excitation power is illustrated with a black horizontal dashed line. (c-f) Spatial density of the synchronised dyad $|\Psi(\mathbf{r},t)|^2$ with size $d=12\;\mathrm{\upmu m}$ calculated at $P_0 = 1.5 P_\text{thr}^{(1)}$. Colour-scale is linear and saturated at 30\% maximum intensity. (g) Spectrally resolved momentum space image of the dyad wavefunction density. Colour-scale is logarithmic. (h) Trajectory of the condensates density imbalance $z$ and relative phase $\phi_1 - \phi_2$ forming a closed loop indicating persistent oscillations with a period $T \approx 10\;\mathrm{ps}$. Vertical dashed line in (a,b) indicates the spectral decomposition of the dyad with $d=12\;\mathrm{\upmu m}$. }
	\label{SupplementaryFigure1}
\end{figure}
\begin{figure}[!ht]
	\center
	\includegraphics[width=0.9\linewidth]{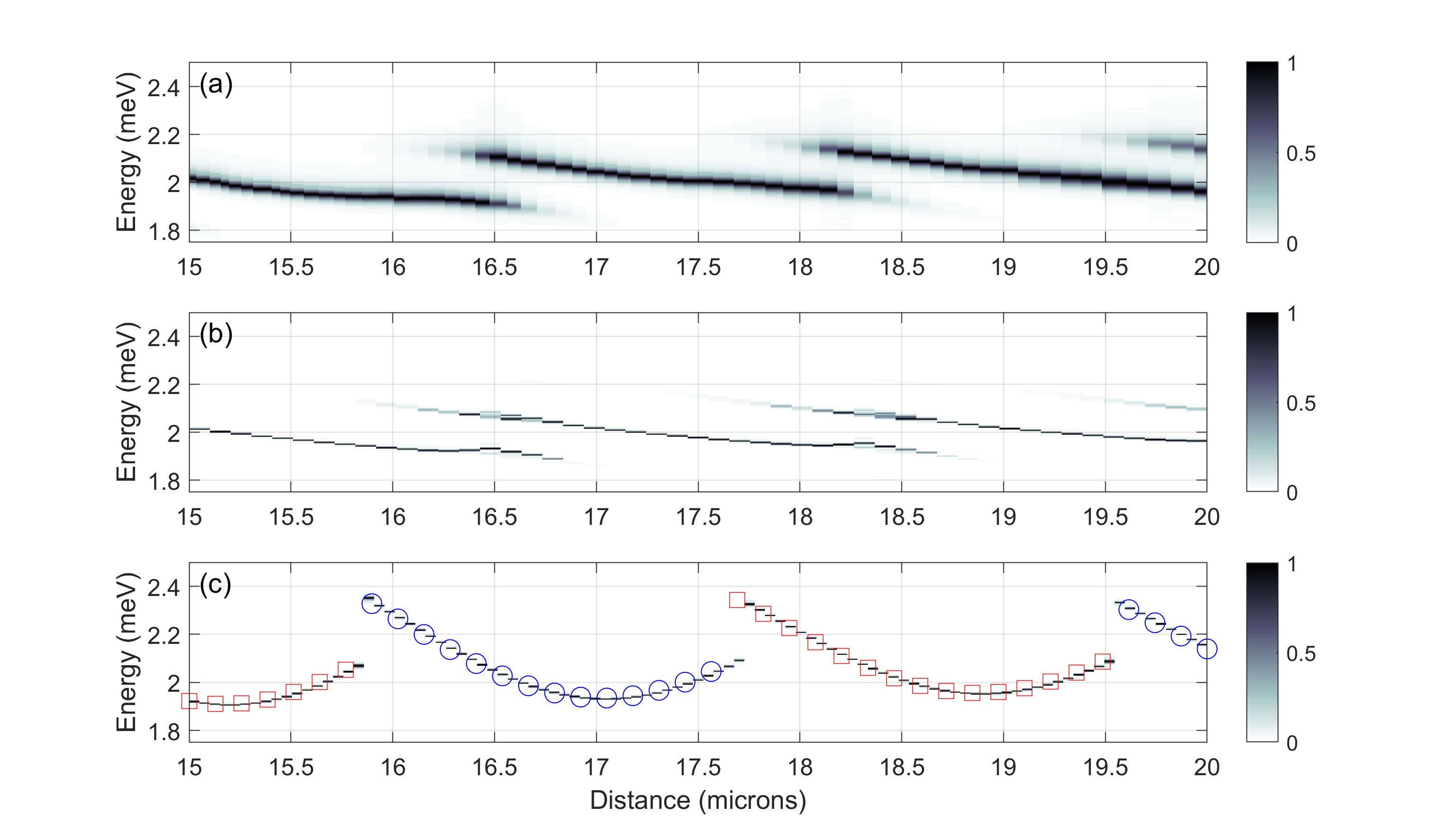}
	\caption{Spectra of two coupled condensates. Comparison between normalised spectral intensity from (a) experiment, (b) Eqs.~(13) and~(14) [main text], and (c) Eqs.~\eqref{eq.CoupledCondensates_7a} and ~\eqref{eq.CoupledCondensates_7b}. Red squares and blue circles denote the only stable energies of in-phase (blue) and antiphase (red) solutions determined from Bogoliubov-de-Gennes stability analysis. Simulation parameters: $\hbar \Omega = (1.22-i 0.5)\; \mathrm{meV}$, $\hbar \alpha = 0.1\;\mathrm{\upmu eV}$, $\hbar R= 0.5\;\mathrm{\upmu eV}$, $\hbar g= 0.5\;\mathrm{\upmu eV}$, $v = 1.9\;\mathrm{\upmu m}\;\mathrm{ps}^{-1}$, $P=100\;\mathrm{ps}^{-1}$, $\Gamma_\mathrm{A}=0.05\;\mathrm{ps}^{-1}$, $\hbar J_0 = 1.1\;\mathrm{meV}$, $k_0=(1.7+i 0.014) \mathrm{\upmu m}^{-1}$ and $\beta = -1$.}
	\label{SupplementaryFigure2}
\end{figure}
\begin{figure}[!ht]
	\center
	\includegraphics[]{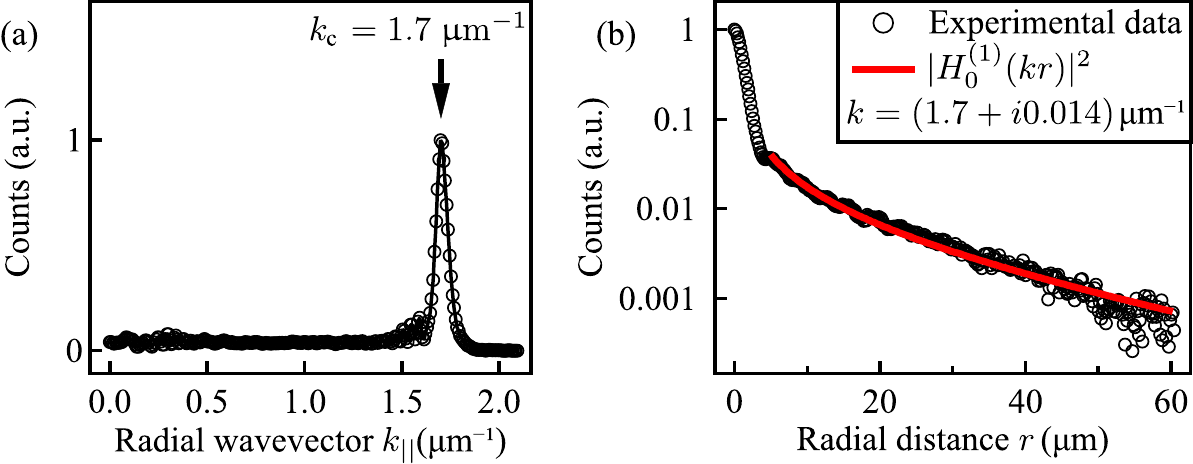}
	\caption{Spatial emission characteristics of a single polariton condensate pumped at $1.5$ times the condensation threshold. (a) Radially-resolved photoluminescence in momentum-space visualises a radial outflow from the condensate centre with well-defined wavevector $k_\mathrm{c}=1.7\;\mathrm{\upmu m}^{-1}$. (b) Measured radial expansion in real-space shows good agreement with a 0-order Hankel-function $H_0^{(1)}(kr)$ for distances $r \geq 5\; \mathrm{\upmu m}$ from the condensate centre.}
	\label{SupplementaryFigure3}
\end{figure}
\begin{figure}[!ht]
	\center
	\includegraphics[]{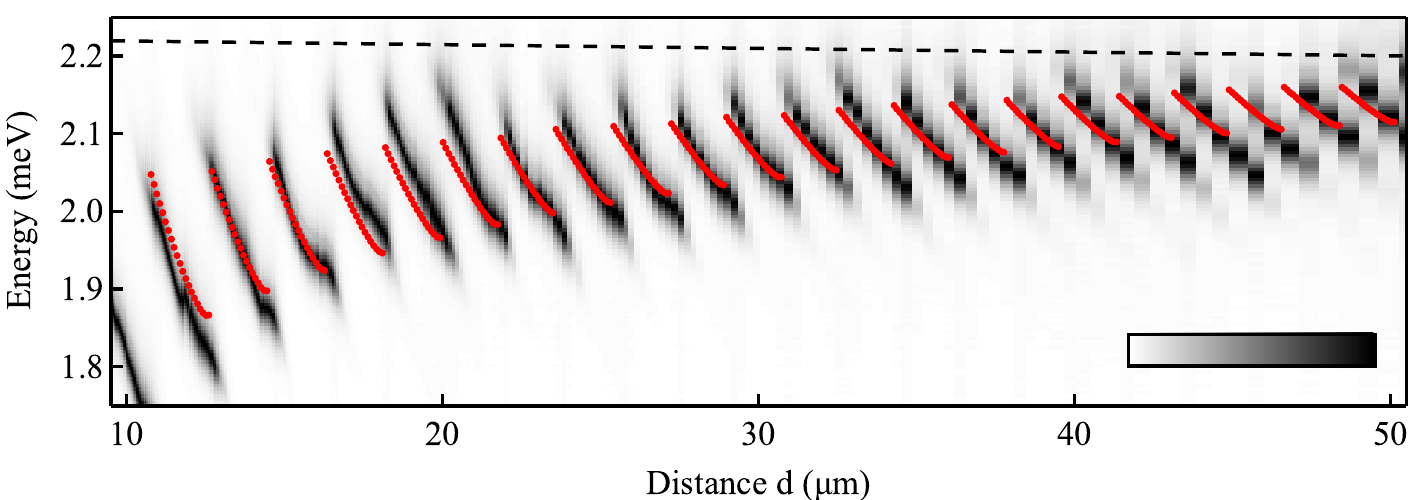}
	\caption{Time-delayed coupled Stuart-Landau oscillators. Comparison of experimentally measured spectra (grey-scale image) of two coupled polariton condensates coupled with separation distance $d$ and numerically calculated fixed-point solutions using a system of two time-delayed coupled Stuart-Landau oscillators (red dots) [see Eq.~(16) in main text]. Parameters: $\hbar \Omega = 2.22\; \mathrm{meV}$, $\hbar \alpha_{\text{eff}} = -0.0229\;\mathrm{meV}$, $\hbar \sigma = 0.0115\;\mathrm{meV}$, $v = 1.9\;\upmu\mathrm{m}\;\mathrm{ps}^{-1}$, $\hbar J_0 = 1.1\;\mathrm{meV}$, $k_0=(1.7+i 0.014) \upmu\mathrm{m}^{-1}$ and $\beta = -1$.}
	\label{SupplementaryFigure4}
\end{figure}
\FloatBarrier 

\section*{Supplementary Note 1: Numerical spatiotemporal simulations}
The dynamics of polariton condensates can be modelled via the mean field theory approach where the condensate order parameter $\Psi(\mathbf{r},t)$ is described by a 2D semiclassical wave equation often referred as the generalised Gross-Pitaevskii equation coupled with an excitonic reservoir which feeds non-condensed particles to the condensate~\cite{Wouters_PRL2007}. The reservoir is divided into two parts: An active reservoir $n_{\mathrm{A}}(\mathbf{r},t)$ belonging to excitons which experience bosonic stimulated scattering into the condensate, and an inactive reservoir $n_{\mathrm{I}}(\mathbf{r},t)$ which sustains the active reservoir~\cite{Lagoudakis2010a, Lagoudakis_PRL2011}.
\begin{align}
i  \frac{\partial \Psi}{\partial t} & = \left[ -\frac{\hbar \nabla^2}{2m} + \frac{g}{2} (n_{\mathrm{A}}+n_{\mathrm{I}}) +  \frac{\alpha}{2} |\Psi|^2 + \frac{i }{2} \left( R n_{\mathrm{A}} - \gamma \right) \right] \Psi,  \label{eq.GPE} \\ 
\frac{ \partial n_{\mathrm{A}}}{\partial t} & = - \left( \Gamma_{\mathrm{A}} + R |\Psi|^2 \right) n_{\mathrm{A}} + W n_{\mathrm{I}}, \label{eq.ResA} \\
\frac{ \partial n_{\mathrm{I}}}{\partial t} & = -  \left(\Gamma_{\mathrm{I}} + W \right) n_{\mathrm{I}} + P(\mathbf{r}). \label{eq.ResI}
\end{align}
Here, $m$ is the effective mass of a polariton in the lower dispersion branch, $\alpha$ is the interaction strength of two polaritons in the condensate, $g$ is the polariton-reservoir interaction strength, $R$ is the rate of stimulated scattering of polaritons into the condensate from the active reservoir, $\gamma$ is the polariton decay rate, $\Gamma_{{\mathrm{A,I}}}$ is the decay rate of active and inactive reservoir excitons respectively, $W$ is the conversion rate between inactive and active reservoir excitons, and $P(\mathbf{r})$ is the non-resonant CW pump profile.

We perform numerical integration of Eqs.~\eqref{eq.GPE}, \eqref{eq.ResA} and~\eqref{eq.ResI} in time using a linear multistep method in time and spectral methods in space. The polariton mass and lifetime are based on the sample properties: $m = 0.28$ meV ps$^2$ $\upmu$m$^{-2}$ and $\gamma = 1/5.5$ ps$^{-1}$. We choose values of interaction strengths typical of InGaAs based systems: $\hbar \alpha = 3.3$ $\upmu$eV  $\upmu$m$^2$, $g = 20 \alpha$. The choice of $g> \alpha$ stems from negatively detuned cavity where polaritons in the condensate are more photonic and thus interact weakly together than those directly with the incoherent reservoir. The non-radiative recombination rate of inactive reservoir excitons is much smaller than the condensate decay rate $\Gamma_{\mathrm{I}}^{-1} = 500$ ps whereas the active reservoir is more ambiguous. It has been argued that the active reservoir (also referred as bottleneck polaritons) decay rate should be larger than the condensate decay rate due to fast thermalisation to the exciton background~\cite{Wouters_2008PRB} and partly because they can decay radiatively. The choice of this value often depends on the type of experiment and ranges over several orders of magnitude~\cite{Lagoudakis_PRL2011, Lagoudakis2010a}. In the current study we choose an intermediate value of $\Gamma_{\mathrm{A}}^{-1} = 20$ ps and find it produces results in good agreement with experiment. It should be noted that the results are not highly sensitive to the exact value of $\Gamma_{\mathrm{A}}$. The final two parameters are then found by fitting experimental results of a single condensate, i.e., the onset of a sharp ring in $k$-space around $k_\mathrm{c} \approx 1.7$ $\upmu$m$^{-1}$ and a blueshift of $\sim 500$ $\upmu$eV when raising the pump power from threshold to twice threshold power $P_\text{thr}^{(1)}$ for a single isolated condensate. The above two measured features of a single condensate are reproduced using the values $\hbar R = 129$ $\upmu$eV $\upmu$m$^{-2}$, and $W = 0.1$ ps$^{-1}$. The pump is written $P(\mathbf{r}) = P_0 e^{-r^2 / 2 w^2}$ where $P_0$ denotes the pump power and $w$ corresponds to a $2\;\mathrm{\upmu m}$ full width at half maximum.

A single numerical scan (i.e., no averaging over many stochastic initial conditions) of the dyad distance dependence is shown in \ref{SupplementaryFigure1}(a-b) and reproduces semi-quantitatively the results of Fig.~3 [main text]. It is worth noting that the current model overestimates the stability of the single energy state in the dyad seen from the plateaus appearing at 0 and 1 in the calculated relative visibility. This can possible stem from the lack of energy relaxation mechanisms, parameter dependence on the polariton Hopfield fraction, and/or overestimation of the active reservoir depletion which can be artificially tuned~\cite{Lagoudakis2010a, Wouters_2010PRB}. Nevertheless, regimes of one-colour and two-colour operation are clearly visible in simulations. All simulations are performed with some natural weak disorder present in the system to account for realistic non-ideal cavity conditions.

In \ref{SupplementaryFigure1}(c-h) we show the condensate wavefunction calculated at $d = 12.7\;\mathrm{\upmu m}$ at $P_0 = 1.5 P_\text{thr}^{(1)}$. Real space dynamics shown in panels (c)-(f) are characterised by periodic beatings in the spatial intensity of the polariton dyad, and the spectrally resolved momentum space image of the wavefunction (g) shows occupancy of two dominant energy levels. In \ref{SupplementaryFigure1}(h) we plot the normalised density imbalance $z = (\rho_1 - \rho_2)/(\rho_1 + \rho_2)$ against the condensate relative phase. Here $\rho_{1,2} = |\Psi(\mathbf{r} - \mathbf{r}_{1,2})|^2$ and $\mathbf{r}_{1,2}$ are the location coordinates of the left and the right pump respectively and $\phi_{1,2} = \arg{[\Psi(\mathbf{r} - \mathbf{r}_{1,2})]}$. The results show a closed trajectory corresponding to persistent oscillations in both phase and particle density with a period $T \approx 10$ ps.

\section*{Supplementary Note 2: Derivation of time delay equation}
Plugging Eq.~(10) [main text] into the time-dependent form of Eq.~(5) [main text],
\begin{equation}
i \hbar \frac{d \Psi(x)}{dt} = \left( -\frac{\hbar^2\partial_x^2}{2m} + V(x)  - i \frac{\hbar \gamma_\mathrm{c}}{2} \right) \Psi(x),
\end{equation}
and integrating out the spatial degrees of freedom gives the following algebraic differential equation,
\begin{equation} \label{eq.SI_CoupledCondensates}
i \hbar  \dot{c}_i  + i \hbar \mathcal{N}_{12}  \dot{c}_j  =   F(c_i,c_j) ,
\end{equation}
where $j=3-i$ ($i=1,2$). The overlap between the two wavefunctions is written,
\begin{equation}
\mathcal{N}_{12}  = e^{-\kappa d} \left( \cos(k_\mathrm{c}d) +\frac{\kappa}{k_\mathrm{c}} \sin(k_\mathrm{c} d)   \right) \simeq e^{-\kappa d} \cos(k_\mathrm{c}d),  
\end{equation}
where $\kappa/k_\mathrm{c} \ll 1$ since the chemical potential of the condensate is much larger than its linewidth. The right-hand side of Eq.~\eqref{eq.SI_CoupledCondensates} is written,
\begin{align} \notag
F(c_i,c_j) & =  \left( \frac{\hbar^2 k^2}{2m} - i\frac{\hbar\gamma_\mathrm{c}}{2} \right) \left(c_i + \mathcal{N}_{12} c_j \right)  +  \left( \left[  V_0 - \frac{i\hbar^2k}{m}  \right] c_i + V_0 e^{ikd} c_j \right) \kappa \\ \label{eq.SI_CoupledCondensates_1}
 & +  \left( \left[  V_0 - \frac{i\hbar^2k}{m}  \right] c_j + V_0 e^{ikd} c_i \right) \kappa e^{-ik^* d}.
\end{align}
We can decouple the above algebraic differential equation by taking into account that the mass matrix,
\begin{equation}
\hat{M} = \begin{pmatrix} 1 & \mathcal{N}_{12} \\ \mathcal{N}_{12} & 1 \end{pmatrix},
\end{equation}
has a well defined inverse since $\mathcal{N}_{12}<1$ for $d>0$. This allows us to write,
\begin{equation}
i \hbar   \begin{pmatrix} \dot{c}_1 \\ \dot{c}_2 \end{pmatrix} =
\hat{M}^{-1} \begin{pmatrix}  F(c_1,c_2) \\  F(c_2,c_1) \end{pmatrix}.
\end{equation}
Assuming that that the resonance condition is satisfied [Eq.(7) in main text],
\begin{equation} \label{Eq.ResonanceCondition}
\left[ \frac{V_0}{\frac{i\hbar ^2 k}{m} - V_0} e^{ikd} \right]^2 = 1,
\end{equation}
and that coupling between condensates is weak, i.e., small $\xi = \exp(- \kappa d)$, one can omit all terms of order $\mathcal{O}(\xi^2)$ and higher. This then gives,
\begin{equation} \label{eq.SI_CoupledCondensates_2}
i\hbar \dot{c}_i  =  \left[ \frac{\hbar^2 k^2}{2m} - i\frac{\hbar\gamma_\mathrm{c}}{2}  + \kappa \left(V_0 -  \frac{i\hbar ^2 k}{m} \right) \right] c_i  +   V_0 \kappa e^{ikd} c_j.  
\end{equation}
It is worth noting that without the resonance condition [Eq.~\eqref{Eq.ResonanceCondition}], instead of arriving at Eq.~\eqref{eq.SI_CoupledCondensates_2} one would get,
\begin{equation} \label{eq.CoupledCondensates_NB}
i \hbar \dot{c}_i =  \left[ \frac{\hbar^2 k^2}{2m} - i\frac{\hbar\gamma_\mathrm{c}}{2} + \kappa \left(V_0 -  \frac{i\hbar ^2 k}{m} \right)  \right] c_i + \left(V_0\cos{(k_\mathrm{c} d)} - \frac{\hbar^2k}{m} \sin{(k_\mathrm{c} d)} \right) e^{-\kappa d} \kappa c_j. 
\end{equation}
Where the inter-condensate coupling term in Eq.~\eqref{eq.CoupledCondensates_NB} is similar to previously established results~\cite{Ohadi2016, Berloff2017a}, but with one critical addition, the complex wavevector $k$ is innately related to the potential 
$V_{0}$.

To illustrate the difference between equations of motion with and without time-delay we replot a portion of the experimental and numerical spectra from Fig.6 [main text] in \ref{SupplementaryFigure2}(a,b). In \ref{SupplementaryFigure2}(c) we then plot the spectra coming from two coupled equations similar to Eqs.~13 and~14 [main text] but but now written:
\begin{align} 
i \dot{c}_i &=  \left[ \Omega  + \left(g + i \frac{R}{2}\right) n_i + \alpha  |c_i|^2  \right] c_i  +  J e^{i\beta } c_j, \label{eq.CoupledCondensates_7a} \\
\dot{n}_i & = -(\Gamma_\mathrm{A} + R |c_i|^2) n_i + P. \label{eq.CoupledCondensates_7b}
\end{align}
Here $J = J_0 H_0^{(1)}(k_0 d)$ is the zeroth order Hankel function of the first kind describing the interference experienced between the two condensates as a function of distance $d$. Such a model was recently theoretically investigated in the context of polariton condensates~\cite{kalinin2019polaritonic} and its relatability to other models in different universality classes, including time-delay models. A major notable feature missing from the no-time-delay model [Eqs.~\eqref{eq.CoupledCondensates_7a} and~\eqref{eq.CoupledCondensates_7b} ] is the overlap between different energy branches indicating cyclical/non-stationary states and the cusp forming at the ends of the branches. Indeed, for small non-linearities ($|c_i|^2 \simeq 0$) Eq.~\eqref{eq.CoupledCondensates_7a} is just a two level model with in-phase and anti-phase eigensolutions $c_{\mathrm{L,U}} = (c_1 \pm c_2)/\sqrt{2}$ whose stability in the non-linear regime ($|c_i|^2 \neq 0$) can be studied through the Bogoliubov-de-Gennes method. Equation~\eqref{eq.CoupledCondensates_7a} also supports non-stationary states at larger $P$ but the spectra qualitatively different from the experimental observations. We therefore believe that time-delay physics as given by Eq.~13 [main text] are an accurate representation of the coupling between ballistically expanding condensates. It describes the effects of interference and how the renormalised self-energy modifies the outflow polariton momentum, giving rise to stable two-colour operation over wide distance intervals.

\section*{Supplementary Note 3: Single condensate}
We characterise the luminescence of a single condensate pumped in the same sample area and with the same pump power, i.e. $1.5$ times the condensation threshold of a single condensate, as the two condensates forming the polariton dyad presented in Fig.~2 and Fig.~3  [main text]. The measured and radially-resolved intensity of far-field and near-field emission are illustrated in \ref{SupplementaryFigure3}(a) and (b), respectively. From (a) we can identify a well-defined outflow-wavevector $k_\mathrm{c} = 1.7\;\mathrm{\upmu m}^{-1}$. Further, by fitting the radial intensity in (b) for $r>5\;\mathrm{\upmu m}$ with the squared magnitude of the 0-order Hankel function of the first kind $|H_0^{(1)}(kr)|^2$ we obtain the complex-valued wavevector $k = (1.7 +i0.014)\;\mathrm{\upmu m}^{-1}$.

\section*{Supplementary Note 4: Time-delayed coupled Stuart-Landau oscillators}
\ref{SupplementaryFigure4} illustrates the numerically calculated brightest fixed point solutions (red dots) of the time-delayed coupled Stuart-Landau oscillators [Eq.~(16) in main text] versus condensate separation distance $d$ matching the experimentally measured and normalised spectra, illustrated in grey-scale. The observed red-shift of the coupled condensate system with respect to the isolated condensate energy level at $2.22\;\mathrm{meV}$ (horizontal dashed line) is the result of an effectively attractive non-linearity, i.e. $\alpha_\text{eff}<0$~\cite{estrecho_single-shot_2018,PhysRevB.89.235310}. Its physical origin is the reduction in reservoir population at each condensate centre due to the inter-condensate stimulated scattering leading to an effectively reduced exciton-polariton interaction energy.

\def\bibsection{\section*{References}} 
\bibliographystyle{naturemag}
\bibliography{refs}

\end{document}